\documentclass[10pt,a4paper,superscriptaddress,aps,prd,showkeys,showpacs,nofootinbib,reprint]{revtex4-2}
\usepackage[utf8]{inputenc}
\usepackage{verbatim}
\usepackage{amsmath,amsfonts,amssymb}
\usepackage[breaklinks,colorlinks]{hyperref}
\usepackage{natbib}
\usepackage{tikz}
\usepackage{float}
\usepackage{graphicx}
\usepackage{adjustbox}
\usepackage{tabularx}
\usepackage{amsbsy}
\usepackage{braket}
\hypersetup{%
,urlcolor=blue
,citecolor=blue
,linkcolor=blue
}

\def\apjl{The Astrophysical Journal Letters}

\begin{document}

\title{Towards Robust Constraints on Axion Dark Matter using PSR J1745-2900}

\author{R.~A.~Battye}
\email[]{richard.battye@manchester.ac.uk}
\affiliation{%
Jodrell Bank Centre for Astrophysics, School of Natural Sciences, Department of Physics and Astronomy, University of Manchester, Manchester, M13 9PL, U.K.
}

\author{J.~Darling}
\email[]{jeremy.darling@colorado.edu}
\affiliation{Center for Astrophysics and Space Astronomy,
Department of Astrophysical and Planetary Sciences,
University of Colorado, 389 UCB
Boulder, CO 80309-0389, USA}

\author{J.~I.~McDonald}
\email[]{jamie.mcdonald@uclouvain.be}
\affiliation{Centre for Cosmology, Particle Physics and Phenomenology,
Université catholique de Louvain,
Chemin du cyclotron 2,
Louvain-la-Neuve B-1348, Belgium}

\author{S.~Srinivasan }
\email[]{sankarshana.srinivasan@postgrad.manchester.ac.uk}
\affiliation{%
Jodrell Bank Centre for Astrophysics, School of Natural Sciences, Department of Physics and Astronomy, University of Manchester, Manchester, M13 9PL, U.K.
}

\label{firstpage}

\date{\today}

\begin{abstract}
We apply novel, recently developed plasma ray-tracing techniques to model the propagation of radio photons produced by axion dark matter in neutron star magnetospheres and combine this with both archival and new data for the galactic centre magnetar PSR J1745-2900. The emission direction to the observer and the magnetic orientation are not constrained for this object leading to parametric uncertainty. Our analysis reveals that ray-tracing greatly reduces the signal sensitivity to this uncertainty, contrary to previous calculations where there was no emission at all in some directions. Based on a Goldreich-Julian model for the 
magnetosphere and a Navarro-Frank-White model for axion density in the galactic centre, we obtain the most robust limits on the axion-photon coupling, to date. These are comparable to those from the CAST solar axion experiment in the mass range $\sim 4.2-60\,\mu{\rm eV}$. If the dark matter density is larger, as might predicted by a ``spike'' model, the limits could be much stronger. The dark matter density in the region of the galactic centre is now the biggest uncertainty in these calculations.
\end{abstract}

\pacs{95.35.+d; 14.80.Mz; 97.60.Jd}

\keywords{Axions; Dark matter; Neutron stars}

\maketitle

{\it Introduction: }Axions have long been known to offer a compelling candidate for dark matter~ \cite{ref:misalign1,ref:misalign2,ref:misalign3} and are motivated by a wide range of models \cite{Arvanitaki:2009fg, Svrcek:2006yi, ref:PQ, ref:K, ref:SVZ, ref:DFSZ, ref:Zhit}. There has recently been a lively theoretical interest in the possibility of axion dark matter detection via its coupling to photons in the magnetospheres of neutron stars~\cite{Pshirkov:2007st,ref:NS-Hook,Leroy:2019ghm,Battye_2020} which relies on resonant conversion at some critical radius $r_{\rm c}$ where the plasma mass, $\omega_{\rm p}=m_{\rm a}$, the axion mass. Radio data have been used to obtain limits on the axion-photon coupling, $g_{{\rm a}\gamma\gamma}$ as a function of $m_{\rm a}$ \cite{Foster:2020pgt,Darling:2020plz,Darling:2020uyo} under the assumption that photon trajectories are radial.

Recent work~\cite{Battye:2021xvt,Witte:2021arp} has shown that this assumption is not sufficiently accurate to predict the signal due to the effects of the plasma and gravity which modify the ray tracing techniques first explored for straight line trajectories in \cite{Leroy:2019ghm}. These techniques allow calculation of the angular, frequency and time-dependence of the signal resulting from the propagation of photons through the magnetosphere. This includes lensing, refraction and frequency distortions, leading, amongst other things, to the Doppler broadening of the line profile~\cite{Battye_2020}.

In this Letter, we combine our novel ray-tracing procedure~\cite{Battye:2021xvt} with new and archival spectral observations of the Galactic Center magnetar PSR~J1745$-$2900, henceforth GCM, using the Karl G. Jansky Very Large Array (VLA\footnote{The National Radio Astronomy Observatory is a facility of the National Science Foundation operated under cooperative agreement by Associated Universities, Inc.}) to search for signatures of axion-photon conversion
in the magnetar magnetosphere.  The new observations are the most sensitive to date, and when combined with
archival VLA observations of Sgr A* already used in \cite{Darling:2020uyo,Darling:2020plz} that include the magnetar in the field of view, provide the strongest
constraints on $g_{{\rm a}\gamma\gamma}$ to date over the mass range $4.2-60\,\mu{\rm eV}$ for a frequency range $\sim 1-15\,{\rm GHz}$.

{\it Modelling the magnetosphere and galactic density: }We use the Goldreich-Julian (GJ) profile \cite{goldreich1969} for number density of the charge carriers, $n_{\rm c}$ in the magnetosphere of the GCM. The parameters describing the model are magnetic field density at the radius of the neutron star, $B_0\approx 1.6\times 10^{14}\,{\rm G}$ \cite{mori2013}, the pulse period, $P\approx 3.76\,{\rm s}$ \citep{kennea2013} which are deduced from fitting to the characteristics of the pulsation, and the mass of the neutron star, $M\approx 1\,M_{\odot}$ assumed to be a similar value to those derived in neutron star models. The object is viewed from position $(\theta,\varphi)$ - $\varphi$ being equivalent to the pulse phase - and the magnetic orientation angle being denoted $\theta_{\rm m}$.

Within the GJ model there is a maximum axion mass for conversion which is set by the critical surface having some part outside the star. After solving for $r_c$ and maximising over $\theta$ and $\varphi$, which involves setting $\theta = \theta_{\rm m}/2$ and $\varphi=0$, this is given by 
\begin{equation}\label{eq:maMAX}
    m_{\rm a}^{\rm max}\approx 85\,\mu{\rm eV}\left({B_0\over 10^{14}\,{\rm G}}\right)^{1\over 2}\left({P\over 1\,{\rm s}}\right)^{-{1\over 2}}\left(1+{1\over 3}\cos\theta_{\rm m}\right)^{1\over 2}\,,
\end{equation}
corresponding to a maximum observing frequency of $\approx 20\,{\rm GHz}$ with the same parametric dependence. For the case of the GCM, where $\theta_{\rm m}$ is not known, this corresponds to a maximum mass $\approx 60\,\mu{\rm eV}$ and a magnetar radius $r_0\approx 10\, {\rm km}$.

The local density of dark matter is assumed to be $\rho_{\rm a}^{\rm local}\approx 0.43\,{\rm GeV}\,{\rm cm}^{-3}$. We can use this to extrapolate to the magnetar's position which is $\approx 0.1\,{\rm pc}$ from the  dark matter peak (assumed to coincide with Sgr A*). In ref. \cite{darling2020apj}, two models were used to make this extrapolation: model A uses an NFW profile with scale radius $18.6\,{\rm kpc}$ \citep{NFW,McMillan2017} to obtain $\rho_{\rm a}=6.5\times 10^4\,{\rm GeV}\,{\rm cm}^{-3}$. Model B has a maximal dark matter spike 
with density $6.4\times 10^{8}\,{\rm GeV}\,{\rm cm}^{-3}$ at the projected magnetar location \citep{Lacroix2018}. Note we also incorporate the local increase in density due to the gravitational field of the pulsar which enhances the density $\rho_a \rightarrow \rho_a \sqrt{2GM/r_c}$ \cite{ref:NS-Hook,Leroy:2019ghm,Battye:2021xvt}. 

{\it Ray tracing vs.~radial trajectories: } Here, we briefly describe our ray-tracing procedure (more details can be found in \cite{Battye:2021xvt}). For each photon trajectory, we solve the following equations which describe the evolution of position 4-vector of the photon $x^\mu = (t,r,\theta, \varphi)$ with respect to a worldline parameter $\lambda$ and its frequency $\omega = \omega(t)$ along rays. They are given by
\begin{align}\label{eq:geodesics1}
&\frac{d^2 x^\mu}{d \lambda^2} + \Gamma^\mu_{\nu \rho} \frac{dx^\nu}{d\lambda} \frac{dx^\rho}{d\lambda} = - \frac{1}{2} \partial^\mu \omega_{\rm p}^2 \,, \\
&\frac{d \omega}{d t} = - \frac{(1-r_s/r)}{2 \omega}\partial_t \omega_{\rm p}^2\,\label{eq:geodesics2},
\end{align}
where $  \omega_{\rm p} = \sqrt{4 \pi \alpha n_c/m_e}$, $m_e$ is the electron mass,  $\alpha = e^2/4\pi$.  Here, $\Gamma_{\nu \rho}^\mu$ are the Christoffel symbols associated to the Schwarzschild metric with Schwarzschild radius $r_{\rm s} = 2GM$. The first of these equations describes the spatial evolution of rays, thereby determining the angular properties of the emission around the pulsar. The second gives the frequency evolution along rays, which allows us to compute not just the signal width, but also make a prediction for the precise line shape of the signal. 

These photons result from the conversion of axions along rays, with each ray, labelled $i$ carrying power $\mathcal{P}_i$ into a solid angle $\Omega_i$ set by
\begin{equation}\label{eq:Ray_Power}
    \frac{d\mathcal{P}_i}{d\Omega_i} =\frac{1}{(n^i_{\rm em})^2 } \tilde{f}(r_{\rm em}, r_{\rm s}) \frac{\rho_{\rm DM}(\textbf{x}^i_{\rm em}) v_{\rm em}^a P_{\rm a \rightarrow \gamma}}{4\pi}\,, 
\end{equation}
where $\rho_{DM}({\bf x}_{\rm em})$ is the dark matter density at the point of emission, $n_{\rm em}$ is the refractive index, $v_{\rm em}^a$ is the axion velocity, and $\tilde{f} (r_{\rm em},r_s)  = 1 - r_{\rm s}/r_{\rm em}  $ is a red-shift factor, all evaluated at the point of emission. The final quantity is the conversion probability \cite{leroy2020}
\begin{equation}\label{eq:Probability}
    P_{\rm a\rightarrow\gamma}  = \frac{\pi g_{a \gamma \gamma}^2 B_{\rm \perp}^2}{2 \omega_{\rm p}'(\textbf{x}_{\rm em}) v^{a}_{\rm em}}\,,
\end{equation}
where $B_{\perp}$ is the magnetic field in the direction perpendicular the photon trajectory. For a given observing direction connecting the observer to the origin at the centre of the pulsar, a plane is constructed perpendicular to the line of sight, with trajectories back-propagated in a normal direction of the plane until they strike the critical conversion surface, where they are assigned a value according to \eqref{eq:Ray_Power} with the total power given by
\begin{equation}
    \frac{d P}{d \Omega} = \sum_i \frac{d P_i}{d \Omega_i}\,.
\end{equation}
Combining eqs.~\eqref{eq:Ray_Power} and \eqref{eq:Probability}, one obtains that the total radiated power scales as $g_{\rm a\gamma\gamma}^2$. One makes use of this fact to put limits on $g_{\rm a\gamma\gamma}$. 



\begin{figure*}[t]
    \centering
    \includegraphics[width = 0.90\textwidth]{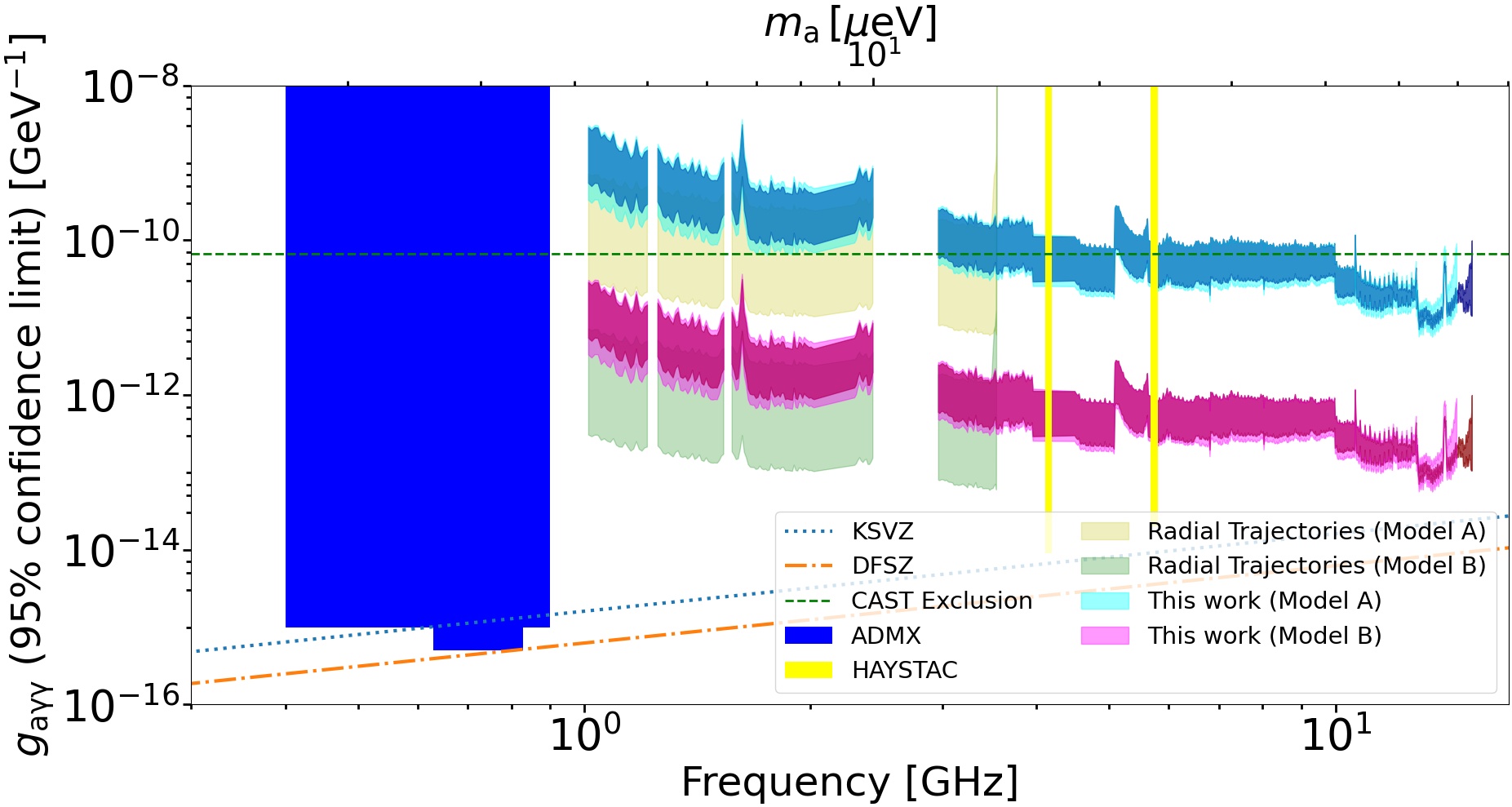}
    \caption{New exclusion limits on axion-dark matter combining archival and recent VLA data and novel plasma ray-tracing techniques \cite{Battye:2021xvt}. The maximum value of $m_a$ for which there is a signal corresponds to eq.~\eqref{eq:maMAX}, beyond which the critical surface disappears entirely inside the star. The thickness of the bands corresponds to the maximum and minimum limits of the signal with respect to the viewing angle $\theta$ and magnetic angle $\theta_m$. The dark coloured bands extremize over $\theta$ for a fixed value of the magnetic angle $\theta_m = 18^\circ$, whilst the lighter bands extremize over both $\theta$ and $\theta_m$.  We also display two possibilities for the axion dark matter density corresponding to models A and B described in the main text. We also show experimental limits from CAST \cite{ref:CAST}, ADMX \cite{ref:ADMX2018} and HAYSTAC \cite{ref:HAYSTAC}. For comparison with our ray tracing analysis, we also display equivalent limits based on a single radial ray connecting the observer to the conversion surface also for $\theta_{\rm m} = 18^\circ$. For the radial formula, above a certain mass cutoff, the signal is zero for some values of $\theta$ and so there is no limit in those cases. For $\theta_m = 18^\circ$, this cutoff occurs at when $m_{\rm a}> 14.5\,\mu{\rm eV}$.}
    \label{fig:g_limits}.
\end{figure*}

{\it Upper limits on flux density in the direction of the GCM: }We have obtained upper limits on the time-averaged spectral line flux density in the direction of the GCM as a function of frequency from three sources. The first two are VLA archival data described in \cite{Darling:2020plz,Darling:2020uyo}.  The third is new VLA observations in its most extended configuration with maximum baseline 36.6~km that provides the angular resolution needed to separate the GCM emission from Sgr A*, which are $\approx 2.4^{\prime\prime}$ apart.
Observations were obtained during seven single-transit sessions between MJD 59184 and 59253 (program 20B-154) in two circular polarizations with 8-bit sampling
and 2~sec recording times.
The new observations are in C-, X-, and Ku-bands (6--8~GHz, 10--12~GHz, and 12--13~GHz) 
and have integration times in the range $1.4-1.6\times 10^{4}\,{\rm sec}$ achieving r.m.s.\ noise levels in 4~MHz channels of 0.13~mJy 
in C-band and $\sim$50--60~$\mu{\rm Jy}$ in the higher frequency bands. 
The data were reduced and calibrated using standard techniques within the CASA package \citep{CASA} following previous work \citep{Darling:2020plz,Darling:2020uyo}.   The continuum emission from Sgr A* is used for interferometric self-calibration (the magnetar continuum is detected but is 2--3 dex fainter than Sgr A*).

Continuum emission was removed by a linear $uv$ subtraction to make null-centered spectral image cubes.
We extracted the GCM spectrum from the cubes using an aperture slightly larger than the synthesized beam and the point source flux density was corrected channel-by-channel. To assess the significance of features in the spectra we form a sky noise spectrum using the r.m.s.\ noise in regions away from the GCM and Sgr A*, and these typically agree with the spectral noise in the GCM spectrum. 

The new and archival GCM and sky spectra were Gaussian smoothed to 4~MHz channels, except for L-band (1--2~GHz), which was 
smoothed to 2~MHz channels.  Noise values range from 0.57~mJy to 52~$\mu$Jy (1.5 and 12.5~GHz).
Unlike the lower-resolution observations presented in \cite{Darling:2020uyo}, no molecular or radio recombination lines were detected in 20B-154. Spatial filtering of the maximum-resolution array removed contaminating signals from the galactic center environment.  

{\it Limits of $g_{{\rm a}\gamma\gamma}$:} No significant ($>4\sigma$) single-channel emission lines were detected at the position of the GCM\footnote{One 5.1$\sigma$ channel at 10.9~GHz was identified with a sidelobe of Sgr A* that was not captured in the sky noise spectrum.} and hence there are no candidate axion signals. These upper limits were then compared with model predictions as a function of $m_{\rm a}$, $\theta$ and $\theta_{\rm m}$ in order to extract 95\% confidence limits on $g_{{\rm a}\gamma\gamma}$. The limits are presented in Fig.~\ref{fig:g_limits} for the two axion densities for halo models A and B described above.  In both cases we have calculated limits for all values of the unconstrained angles $\theta$ and $\theta_{\rm m}$ and presented the range. We see that for halo model A the limits are similar to those from the CAST solar axion experiment \cite{ref:CAST}, while they are a couple of orders of magnitude stronger (the limit on $g_{{\rm a}\gamma\gamma}$ is $\propto \sqrt{\rho_{\rm a}}$) for model B.

It is interesting to contrast this state-of-the-art treatment with the original toy-setup considered in ref.~\cite{ref:NS-Hook} that only considered radially outgoing photons where the angular dependence factorises and 
\begin{align}\label{eq:hookPower}
\frac{dP^{\rm radial}}{d\Omega}\left(\theta, \theta_{\rm m}, \varphi\right) \propto
   \frac{3\, \left(\hat{\textbf{m}}\cdot\hat{\textbf{r}}\right)^2 + 1}{\left|3 \cos \theta \, \hat{\textbf{m}}\cdot \hat{\textbf{r}} - \cos \theta_m \right|^{4/3}}\,,
\end{align}
where $\textbf{m}$ is the magnetic dipole and hats denote unit vectors. We have also presented limits for these radial trajectories in Fig.~\ref{fig:g_limits}. These are somewhat stronger than one gets from ray tracing, and also cut-off at a lower axion mass, $m_{\rm a}\gtrsim 14\,\mu{\rm eV}$, due to the fact that beyond that there are points in the $\theta-\theta_{\rm m}$ plane where there is no predicted signal.
 
{\it Conclusions:} Our principle conclusion is that robust constraints on axion dark matter cannot be obtained using simplistic radial trajectories ~\cite{ref:NS-Hook,Foster:2020pgt} and therefore ray-tracing \cite{Leroy:2019ghm,Witte:2021arp,Battye:2021xvt} becomes unavoidable. The principle reason for this is that radial trajectories connect each viewing angle in the sky with a single point on the emission surface resulting in a very undemocratic spread of power across the sky. This leads to very sharp angular variation in the signal with a few narrow bright-spots that are associated to trajectories with high emission rates and for higher masses, so much angular variation that for many viewing angles the pulse averaged power vanishes entirely. As a result, in the radial approximation, one cannot be sure a signal is even present. 

In future work it may be possible by modelling the pulsar beam and fitting to radio observations of the pulse width, one could infer something about $\theta_{\rm m}$, and perhaps the viewing angle relative to the beam axis $\theta - \theta_{\rm m}$, this may allow us to attach some probability distribution to $\theta$ and $\theta_{\rm m}$, reducing the parametric uncertainty. Nonetheless, at present the most conservative approach is to assume the viewing and magnetic angles are unknown parameters, and take those values $(\theta_{\rm m}, \theta)$ for which the pulse-averaged power is minimal. With this in mind, the radial-trajectory approximations of \cite{ref:NS-Hook} clearly make it impossible to place reliable bounds since the signal entirely vanishes for many viewing angles at higher masses. 

Rather pleasingly, our analysis also reveals that the angular variation in power is actually under good control, typically around an order of magnitude for the values chosen in this paper. This translates to less than an order of magnitude uncertainty in the sensitivity to $g_{{\rm a}\gamma \gamma}$ which scales as the square root of the power. 

To summarise, one can identify three principle uncertainties in axion-constraints. (i) The dark matter profile and therefore the dark matter density near the neutron star.  (ii) The sensitivity of axion limits to the structure of the magnetosphere in terms of the magnetic field and plasma density profiles. (iii) The angular, frequency and time-dependence of the signal which depends on the propagation of photons through the magnetosphere. 
Point (i) is of course a challenge for any attempt to detect dark matter, even here on earth where the velocity and density is not necessarily known. It can also be that this uncertainty can be reduced by choosing pulsar further from the galactic centre where there is less disagreement between different halo models. Future observations may also allow better determination of the dark matter density around the magnetar. However, at the moment this is the primary uncertainty. It may be possible to ameliorate point (ii) in future work by applying state-of-the-art magnetosphere modelling and examining the variation in signal properties across different models. The subject of this letter is point (iii), which has recently been resolved by sophisticated ray-tracing techniques as we have shown. In addition it may be necessary to include the full range of effects considered in refs.~\cite{Witte:2021arp} and \cite{Battye_2020}.

{\it Acknowledgements: }We are grateful to Samuel Witte for discussions on plasma ray tracing. JIM is supported by the F.R.S.-FNRS under the Excellence of Science (EOS) project No.~30820817 (be.h). SS is supported by a George Rigg Scholarship from the University of Manchester.  We thank the VLA operations, observing, and computing staff who made this work possible.  This research made use of CASA \citep{CASA}, {\tt NumPy} \citep{NumPy}, {\tt Matplotlib} \citep{Matplotlib}, and {\tt Astropy} \footnote{http://www.astropy.org}, a community-developed core Python package for Astronomy \citep{astropy:2013, astropy:2018}.

\bibliographystyle{apsrev4-1}
\bibliography{Ref.bib}

\end{document}